\documentclass{PoS}

\usepackage{amsmath}
\usepackage{amssymb}
\usepackage{mathrsfs}
\usepackage{amsfonts}
\usepackage{soul}

\def\be{\begin{equation}}
\def\ee{\end{equation}}
\newcommand{\bea}{\begin{eqnarray}}
\newcommand{\eea}{\end{eqnarray}}

\usepackage{cite}
\usepackage{xfrac} 
\usepackage{fontawesome} 
\usepackage{booktabs} 
\def\deltae{\psi_e}
\def\deltamu{\psi_\mu}
\def\one{\faThumbsOUp}
\def\zero{\faThumbsDown}
\def\tabcolor{black}

\title{Interference effects in semileptonic decays from heavy Majorana neutrinos}

\ShortTitle{Interference effects from heavy Majorana neutrinos}

\author{\speaker{X. Marcano}$^{\,a}$, A. Abada$^{a}$, C. Hati$^{b}$ and A. M. Teixeira$^{b}$\thanks{Work supported by the European Union’s Horizon 2020 research and innovation programme under the Marie Skłodowska-Curie grant agreements No 690575 and No 674896.}\\
       $^{a}$ Laboratoire de Physique Th\'eorique, CNRS\\
Univ. Paris-Sud, Universit\'e Paris-Saclay, 91405 Orsay, France\\
$^{b}$ Laboratoire de Physique de Clermont (UMR 6533), CNRS/IN2P3,\\ 
Univ. Clermont Auvergne, 4 Av. Blaise Pascal, F-63178 Aubi\`ere Cedex, France\\
        E-mail: \email{xabier.marcano@th.u-psud.fr},  \email{ asmaa.abada@th.u-psud.fr}, \email{chandan.hati@clermont.in2p3.fr}, \email{ana.teixeira@clermont.in2p3.fr}}       

\abstract{
Several beyond the Standard Model scenarios introduce new heavy neutrinos, whose Dirac or Majorana nature could be tested by comparing the rates of lepton number violating and lepton number conserving processes: a Dirac fermion induces only the latter, while a Majorana one predicts the same rate for both of them. Nevertheless, in the presence of more than one Majorana fermion, this picture may change drastically due to interference effects. 
We focus on lepton number violating and lepton flavour violating semileptonic meson decays induced by two heavy Majorana fermions, exploring the necessary conditions to have sizeable interference effects and discussing their implications for current experimental constraints and possible future observations. In particular, we show how the $CP$ violating phases may lead to  an enhancement of the lepton number violating modes and suppression of the lepton number conserving ones, and vice-versa.
}

\FullConference{
European Physical Society Conference on High Energy Physics - EPS-HEP2019 -\\
			10-17 July, 2019\\
			Ghent, Belgium}

\begin{document}

\section{Introduction}

In many extensions of the Standard Model (SM) new neutral leptons are introduced in order to explain the observed oscillation phenomena. 
These new states would not be charged under the SM gauge group, motivating the name of sterile neutrinos, and making their discovery a very challenging task. 
An interesting experimental signature to probe the hypothesis of the existence of these particles is that of lepton number violating (LNV) processes, since these would not only reveal the existence of New Physics, but also the Majorana nature of the exchanged fermions.

We focus  on the LNV meson semileptonic and tau three-body decays, as well as their lepton number conserving (LNC) counterparts. 
A particular example could be the semileptonic kaon decays $K\to\pi \ell\ell'$ that are being currently searched for in NA62~\cite{CortinaGil:2019dnd}.
These processes can be induced by the exchange of a sterile neutrino, whose contribution could be resonantly enhanced if the neutrino could be produced on-shell. 
Here, we extend previous studies to the case with two  sterile neutrinos and explore the importance of the interference effects between their contributions.

In the presence of a single sterile neutrino, the same branching ratios are predicted for both LNV and LNC decays if the neutrino is of Majorana nature, whereas only LNC processes can take place when dealing with a Dirac neutrino.
As we will see, in the case with two Majorana neutrinos, the interference between them may play an important role, leading to strong  interferences in either the LNV or the LNC processes.
We will also discuss how the complementarity between different LNV and LNC channels may help disentangle  the single-neutrino hypothesis from the interfering-neutrinos one.
For further discussion and details on the study, we refer to our main work~\cite{Abada:2019bac}. 

\section{Semileptonic meson decays from resonant sterile neutrinos}

We consider the framework of simplified SM extensions via the addition of $N=2$ extra sterile neutrinos, which interact with the SM particles only via their mixings $U_{\alpha i}$ to the light active neutrinos.
Together with the heavy neutrino masses $m_{i}$, these mixings are the relevant parameters for the forthcoming analysis.
We find it useful to denote them in terms of the modulus and phases,
\be\label{angle-phase}
U_{\alpha i}=e^{-i\phi_{\alpha i}}|U_{\alpha i}|, \quad
\alpha=e,\mu,\tau, \text{ and }  i=4,5\,. 
\ee
Due to their mixings, the sterile neutrinos may induce LNC  semileptonic processes $M\to M' \ell_\alpha^\pm \ell_\beta^\mp$  and the corresponding LNV ones $M\to M' \ell_\alpha^\pm \ell_\beta^\pm$, $M$ and $M'$ being pseudoscalar mesons.
Assuming that the dominant contributions are those from the sterile neutrinos,  the corresponding squared amplitudes depend on the neutrino parameters as 
\begin{align}
\left|\mathcal{A}^{\rm LNC}_{M\to M' \ell_\alpha^\pm
  \ell_\beta^\mp}\right|^2 \!\!&\propto 
 \big|U_{\alpha 4}\big|^2 \big|U_{\beta4}\big|^2
|g(M)|^2\, \Big|1+\kappa'\, e^{\mp i(\psi_{\alpha
  }-\psi_{\beta})}\Big|^2\,,\label{GamamLNC}\\ 
\left|\mathcal{A}^{\rm LNV}_{M\to M' \ell_\alpha^\pm
  \ell_\beta^\pm}\right|^2 \!\!&\propto 
 \!\big|U_{\alpha 4}\big|^2
\big|U_{\beta4}\big|^2 |f(M)|^2\, \Big|1+\kappa\, e^{\mp
  i(\psi_{\alpha}+\psi_{\beta})}\Big|^2 , \label{GamamLNV} 
\end{align}
where we have defined the relative phases $\psi_\alpha\equiv \phi_{\alpha_5}-\phi_{\alpha4}$, the average mass $M$ and the mass splitting $\Delta M$ of the two sterile neutrinos. 
The functions  $f$ and $g$ are the integrals one obtains when
computing the decay amplitudes for LNV and LNC semileptonic decays of
mesons  (details can be found for instance in \cite{Abada:2017jjx}).
The complex quantities $\kappa$ and $\kappa'$ reflect the
relative size of the contributions of the two sterile fermions to the
processes
\begin{equation}\label{kappas}
\kappa\equiv\dfrac{ |U_{\alpha5} U_{\beta5}| }{|U_{\alpha4}
  U_{\beta4}|}\dfrac{f(m_5)}{f(m_4)}, \quad \kappa'\equiv\dfrac{
  |U_{\alpha5} U^*_{\beta5}| }{|U_{\alpha4}
  U^*_{\beta4}|}\dfrac{g(m_5)}{g(m_4)} \,.
\end{equation} 

From these simple equations we can already see that the LNC and LNV amplitudes have orthogonal dependencies with respect to the relative phases, i.e., they depend on $\psi_{\alpha}-\psi_{\beta}$ and $\psi_{\alpha}+\psi_{\beta}$, respectively.
This shows the complementarity between the two channels, which is important when exploring possible interference effects, as we will discuss later. 
Moreover, these combinations of relative phases reveal the origin of each of the $CP$ phases, as any Majorana phase should cancel in the $\psi_{\alpha}-\psi_{\beta}$ combination of the LNC process~\cite{Abada:2019bac}.

In order to discuss the impact of the interference on the LNV and LNC decay amplitudes, we consider the quantities $R_{\ell_\alpha\ell_\beta}$ and $\widetilde R_{\ell_\alpha\ell_\beta}$, defined as 
\begin{equation}\label{ratio}
R_{\ell_\alpha\ell_\beta} \equiv
\frac{\Gamma^{\rm LNV}_{M\to M' \ell_\alpha^\pm
    \ell_\beta^\pm}}{\Gamma^{\rm LNC}_{M\to M'
    \ell_\alpha^\pm \ell_\beta^\mp}} \,, 
\qquad
\widetilde R_{\ell_\alpha\ell_\beta} \equiv
\frac{\Gamma^{\rm LNC}_{M\to M' \ell_\alpha^\pm
    \ell_\beta^\mp}-\Gamma^{\rm LNV}_{M\to M'
    \ell_\alpha^\pm \ell_\beta^\pm}} 
{\Gamma^{\rm LNC}_{M\to M' \ell_\alpha^\pm
    \ell_\beta^\mp}+\Gamma^{\rm LNV}_{M\to M'
    \ell_\alpha^\pm \ell_\beta^\pm}} 
=\frac{1-R_{\ell_\alpha \ell_\beta}}{1+R_{\ell_\alpha \ell_\beta}}\,.    
\end{equation}
The second ratio, $\widetilde R_{\ell_\alpha\ell_\beta}$, is related to the usually considered $R_{\ell_\alpha\ell_\beta}$; nevertheless it will be useful to understand the interference effects in a more general situation, as it is well-defined even if there is a strong suppression in the LNC channel.

\section{Interference effects from two Majorana neutrinos}

The contributions from the sterile neutrinos to the semileptonic meson decays become dominant in the mass range where the neutrinos can be produced on-shell, in which case  one can have a resonant enhancement.
Assuming the narrow-width approximation where the sterile neutrino width $\Gamma_{N_i}$ is much smaller than its mass $m_i$, this  resonant enhancement can be understood as an increase of $\mathcal O(m_i/\Gamma_{N_i})$ in the decay rates.

In the case of the SM extended by only one heavy Majorana neutrino in this resonant regime, the predictions for the LNV and LNC decay widths are equal, implying that $R_{\ell_\alpha \ell_\beta}=1$ and thus $\widetilde R_{\ell_\alpha\ell_\beta} =0$. 
Nevertheless, in the presence of two (or more) Majorana neutrinos in the resonant regime, the interference effects may change the predictions for  $R_{\ell_\alpha \ell_\beta}$ and $\widetilde R_{\ell_\alpha\ell_\beta}$.
In particular, this will happen when the mass splitting of the heavy Majorana states is very small, $\Delta M < \Gamma_N$, since the overlap  between their  contributions may lead to destructive or constructive interferences.
Moreover, in order to have sizeable interference effects, the relative size of the contributions of the two neutrinos to each amplitude should be of the same order, implying that they should mix with similar strength to the relevant active flavours.

All these conditions for maximal interference effects can be summarized in terms of the $\kappa^{(}{'}^{)}$ factors and the ratio $R_{\ell_\alpha\ell_\beta}$.
The former can be expanded as follows,
\begin{equation}\label{kappa}
|\kappa|\simeq |\kappa'|
= \dfrac{ |U_{\alpha5} U^*_{\beta5}| }{|U_{\alpha4} U^*_{\beta4}|}
\Big(1+{\mathcal{O}} \Big( \frac{\Delta M}{\Gamma_N} \Big)\Big)\ ,
\end{equation} 
while for the ratio $R_{\ell_\alpha\ell_\beta}$ we have,
\begin{equation}\label{ratio:simple-gen}
R_{\ell_\alpha\ell_\beta}=
\frac{(1- |\kappa|)^2 + 4 |\kappa|
  \cos^2\left({\delta\pm(\psi_\alpha+\psi_\beta)\over 2}\right)}{(1-
  |\kappa'|)^2 + 4 |\kappa'|
  \cos^2\left({\delta'\pm(\psi_\alpha-\psi_\beta)\over 2}\right)} \,, 
\end{equation}
where we have set $\kappa^{(}{'}^{)}= |\kappa^{(}{'}^{)}|e^{ i \delta^{(}{'}^{)}}$, and with the $\pm$ referring to the electric charge of the lepton $\alpha$. 
If the mass splitting is very large or one neutrino contributes dominantly, i.e.  very small or very large $|\kappa^{(}{'}^{)}|$, the interference between the two neutrinos is not relevant and the single Majorana neutrino prediction, $R_{\ell_\alpha\ell_\beta}=1$, is recovered. 
On the other hand, if $|\kappa|\sim|\kappa'|\approx 1$, the interference becomes relevant and $R_{\ell_\alpha\ell_\beta}$ may deviate from unity.  

\begin{figure}[t!]
\begin{center}
\includegraphics[width=.5\textwidth]{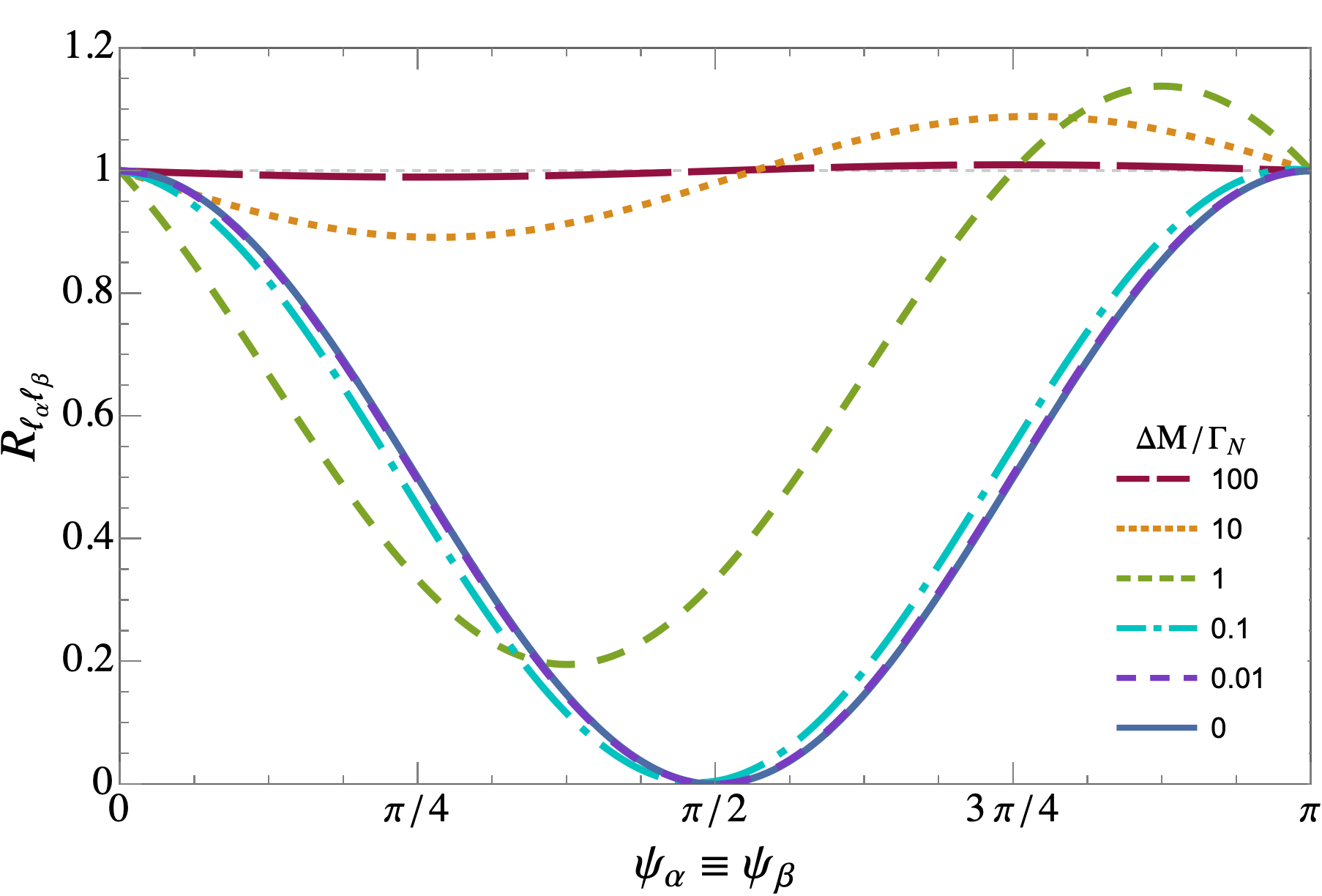}
\caption{
The ratio $R_{\ell_\alpha\ell_\beta}$  as a function of a common relative $CP$  violating phase. Several regimes of $\Delta M/\Gamma_N$ illustrate how the conditions discussed in the text are crucial to observe maximal interference effects.     
}\label{plotRonephase}
\end{center}
\end{figure}

We illustrate in Fig.~\ref{plotRonephase} the behaviour of the ratio $R_{\ell_\alpha\ell_\beta}$ in the simplified case $\psi_\alpha=\psi_\beta$, as a function of this common relative $CP$ phase, and  for different values of $\Delta M/\Gamma_N$.
The lines of Fig.~\ref{plotRonephase} were obtained via a full numerical evaluation~\cite{Abada:2019bac}, although we  see that they follow the above qualitative discussion.  
Notice that in this simplified case only the LNV is sensitive to the relative phase, and that it suffers both constructive and destructive interference effects, clearly signaling the presence of at least two additional sterile states. 
This is most prominent  in the case of $\Delta M/\Gamma_N \to 0$, where very large suppressions happen if the sterile neutrinos have opposite phases. 
Indeed, this is the case in models where  lepton number conservation is imposed. 
On the other hand, the prediction of a single Majorana neutrino hypothesis, $R_{\ell_\alpha\ell_\beta}=1$, is recovered for vanishing values of the relative phase, when the interference is constructive for both LNC and LNV channels, or in the case of $\Delta M/\Gamma_N \gg 1$, since the interference effects become negligible and the contribution of the two sterile neutrinos can be added incoherently. 

The  interference effects illustrated in  Fig.~\ref{plotRonephase} are a general feature of semileptonic meson and tau decays.
Moreover, the same interference pattern is obtained for the ratio of same-sign to 
opposite-sign dilepton number of events in hadron colliders~\cite{Das:2017hmg}.
Nevertheless, in a more general case where $\psi_\alpha\ne\psi_\beta$, destructive interferences can occur in both LNV and LNC decay amplitudes, leading in the latter case to enhanced values of $R_{\ell_\alpha\ell_\beta}$. 
We stress, however, that this kind of suppression can only occur in the LNC processes if the final state charged leptons have different flavour, as the relative $CP$ phases cancel out in the same flavour case. 

\begin{figure}[t!]
\begin{center}
\includegraphics[width=.55\textwidth]{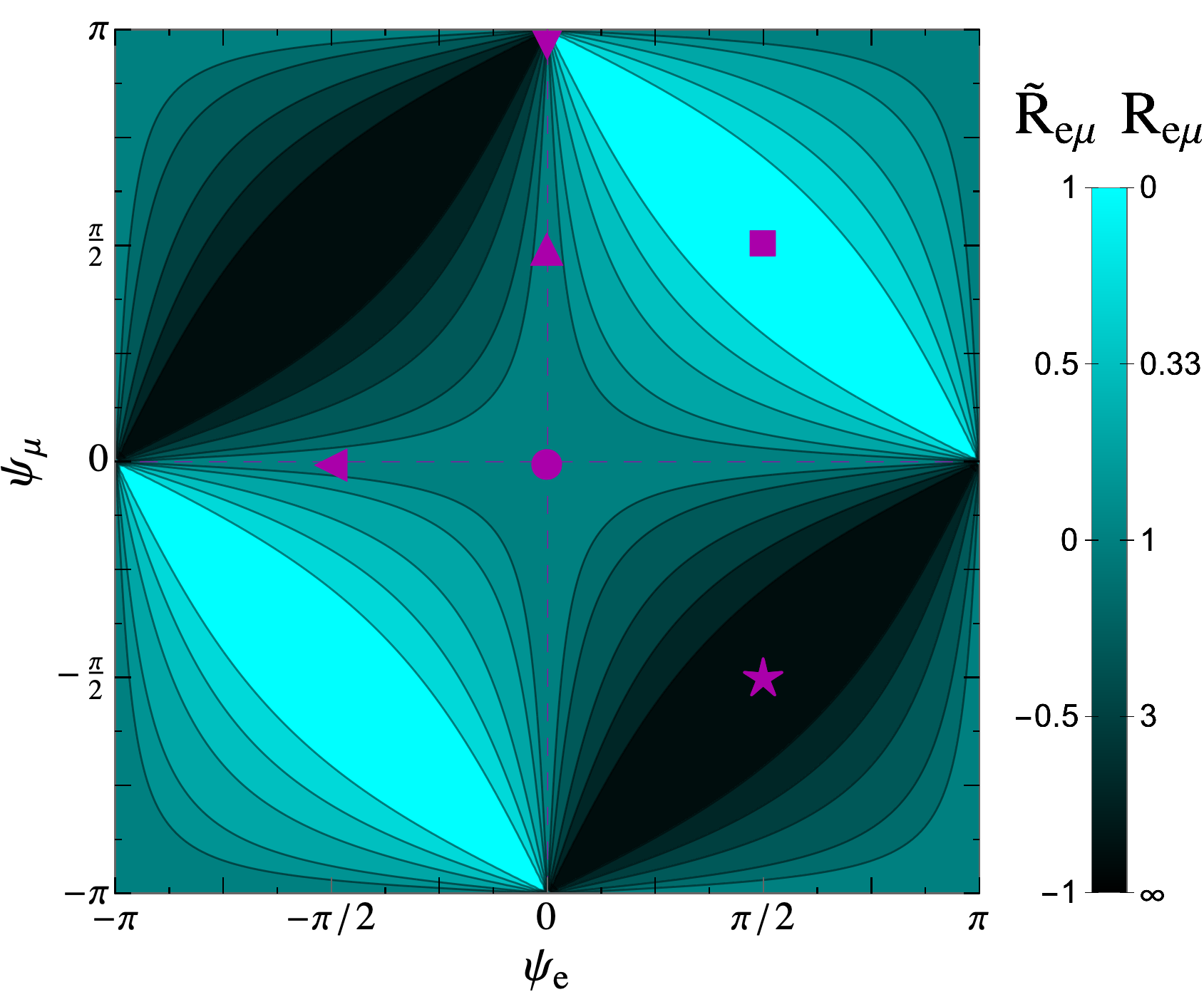}
\caption{
Ratios $R_{e\mu}$  and $\widetilde R_{e \mu}$  of the LNV and LNC decays $K\to\pi e \mu$.
Details on the neutrino parameters can be found in~\cite{Abada:2019bac}.
 Symbols are benchmark points for the discussion, see Table~\ref{tableBenchmarks}.
}\label{plotRprimeemu}
\end{center}
\end{figure}

We display in Fig.~\ref{plotRprimeemu} the ratios $R_{e\mu}$ and $\widetilde R_{e\mu}$ on the parameter space spanned by the two relative phases,  $\psi_e$ and $\psi_\mu$.
As before, the numerical computation has been done for a particular benchmark point (with $\kappa=\kappa'=1$ to maximize the interference), and for the decay channel $K\to\pi e\mu$.
Nevertheless similar results are obtained for other semileptonic meson and tau decays, provided that the neutrinos are in the resonant regime and that the leptons have different flavour. 

Due to the above discussed orthogonal  dependence on the relative phases for the LNV and LNC decays,  the cancellation leading to the extreme case of vanishing  LNC amplitudes corresponds in some cases to maximal values for the LNV, and vice-versa. 
This leads to the bright regions in Fig.~\ref{plotRprimeemu} where the LNV channel is suppressed, $R_{e\mu}\ll1$, and to the dark regions with a suppressed LNC decay, $R_{e\mu}\gg1$.

These results  clearly illustrate the role of the interferences regarding the potential observation of each transition, and strongly suggest  that any conclusion regarding the contribution of sterile fermions to LNV semileptonic meson decays must be accompanied by the study of the corresponding (flavour violating) LNC mode. 
Even if a combination of phases leads to an experimentally ``blind spot''  in which the LNV ratio lies beyond sensitivity due to destructive interference effects, the same interference might be constructive for the corresponding LNC mode (and vice-versa).  

This discussion can be extended to include also the LNV decay modes leading to same-flavoured final state leptons, which are also sensitive to the relative $CP$ violating phases as can be seen in Eq.~\eqref{GamamLNV}.
By studying all these channels and comparing their relative size with respect to what we could expect in the single sterile neutrino hypothesis, we could infer the presence of at least two interfering neutrinos. 

\begin{table}[t!]
\begin{center}
\begin{tabular}{l|c|cccc}
& $(\deltae,\deltamu)$ & $e^\pm e^\pm$ & $\mu^\pm \mu^\pm$ & $e^\pm \mu^\pm$ & $ e^\mp \mu^\pm$ \\[.2ex]
\hline
&  & &&&\\[-2.2ex]
{\color{\tabcolor}\Large $\bullet$} 
& $(0, 0)$ & \one &\one &\one & \one\\
{\color{\tabcolor}\small $\blacksquare$}
& $(\sfrac\pi2, \sfrac\pi2)$ & \zero &\zero &\zero & \one\\
{\color{\tabcolor}\small $\bigstar$ }
& $(\sfrac\pi2, \sfrac{-\pi}2)$ & \zero &\zero &\one & \zero\\
{\color{\tabcolor} $\blacktriangledown$}
& $(0, \pi)$ & \one &\one &\zero & \zero\\
{\color{\tabcolor} $\blacktriangle$}
& $(0, \sfrac\pi2)$ & \one &\zero &  \one/2 &  \one/2 \\
{\color{\tabcolor} $\blacktriangleleft$  }
& $(-\sfrac\pi2,0)$ & \zero &\one &  \one/2 &  \one/2 \\
\end{tabular}
\caption{
Qualitative representation of the interference effects in $M\to M' \ell_1\ell_2$.
The symbol \one~ reflects a constructive interference,  \zero~ a destructive one, and \one/2 means an intermediate case.
The symbols in the first column correspond to the benchmark points in Fig.~\ref{plotRprimeemu}. }\label{tableBenchmarks}
\end{center}
\end{table} 

We schematically illustrate this idea in Table~\ref{tableBenchmarks}, where we qualitatively sketch the interference behaviour for four complementary decay channels (with final state $e^\pm e^\pm$, $\mu^\pm \mu^\pm$, $e^\pm \mu^\pm$ and $ e^\mp \mu^\pm$), and for the benchmark points marked with purple symbols in Fig.~\ref{plotRprimeemu}.
The first row  corresponds to a case of two degenerate sterile neutrinos with the same $CP$ phases, so they interfere constructively in all channels. 
Under this hypothesis, the predictions in terms of ratios such $R_{e\mu}$ are the same as those in the single neutrino hypothesis, and consequently we cannot disentangle between having only one Majorana neutrino or more. 
In the second row, the phases are such that all LNV processes are suppressed, $R_{e\mu}\ll0$, as it happens in low scale seesaw models imposing a lepton number symmetry.
Notice however that this experimental signature would be that of a single Dirac neutrino.
On the other hand, the other rows show scenarios that could not be obtained with a single sterile neutrino,  Dirac or Majorana.
Consequently, if one of these latter patterns is to be observed, it would point towards the existence of two, or more, interfering sterile Majorana neutrinos. 

\section{Conclusions}

We have studied the impact of constructive and destructive interference effects on the contributions of sterile Majorana fermions to the decay rates of lepton number conserving and lepton number violating semileptonic  decays. 
Provided that some of these semileptonic decays would be experimentally observed, we have highlighted that the study of the different LNC and LNV channels would help disentangling the single neutrino hypothesis from the one with several interfering neutrinos. 
On the other hand, given the current negative results in the searches for these processes, we remark that the interference effects discussed here should be taken into account for a proper re-interpretation of the results in terms of more than one sterile neutrino.

%
\bibliographystyle{JHEP}
 \bibliography{biblio}

\providecommand{\href}[2]{#2}\begingroup\raggedright\begin{thebibliography}{1}

\bibitem{CortinaGil:2019dnd}
{\scshape NA62} collaboration, \emph{{Searches for lepton number violating
  $K^+$ decays}},
  \href{https://doi.org/10.1016/j.physletb.2019.07.041}{\emph{Phys. Lett.}
  {\bfseries B797} (2019) 134794}
  [\href{https://arxiv.org/abs/1905.07770}{{\ttfamily 1905.07770}}].

\bibitem{Abada:2019bac}
A.~Abada, C.~Hati, X.~Marcano and A.~M. Teixeira, \emph{{Interference effects
  in LNV and LFV semileptonic decays: the Majorana hypothesis}},
  \href{https://doi.org/10.1007/JHEP09(2019)017}{\emph{JHEP} {\bfseries 09}
  (2019) 017} [\href{https://arxiv.org/abs/1904.05367}{{\ttfamily
  1904.05367}}].

\bibitem{Abada:2017jjx}
{A. Abada {\it et al.}}, \emph{{Effective Majorana mass matrix from tau and
  pseudoscalar meson lepton number violating decays}},
  \href{https://doi.org/10.1007/JHEP02(2018)169}{\emph{JHEP} {\bfseries 02}
  (2018) 169} [\href{https://arxiv.org/abs/1712.03984}{{\ttfamily
  1712.03984}}].

\bibitem{Das:2017hmg}
A.~Das, P.~S.~B. Dev and R.~N. Mohapatra, \emph{{Same Sign versus Opposite Sign
  Dileptons as a Probe of Low Scale Seesaw Mechanisms}},
  \href{https://doi.org/10.1103/PhysRevD.97.015018}{\emph{Phys. Rev.}
  {\bfseries D97} (2018) 015018}
  [\href{https://arxiv.org/abs/1709.06553}{{\ttfamily 1709.06553}}].

\end{thebibliography}\endgroup
          
\end{document}